\documentclass[runningheads]{llncs}
\usepackage{hyperref}
\usepackage{tikz}
\usepackage{subcaption}
\usepackage{algpseudocode}
\usepackage{algorithm}
\usepackage{stackengine}
\usepackage{amssymb}
\usepackage{caption}
\usepackage{amsmath}
\usepackage{booktabs}
\usepackage{bold-extra}
\usepackage{float}
\usepackage{bm}

\DeclareCaptionLabelFormat{andtable}{#1~#2  \&  \tablename~\thetable}

\algrenewcommand{\alglinenumber}[1]{\tiny#1:}

\newcommand{\textunderset}[2]{\begin{tabular}[t]{@{}c@{}}#2\\[-0.3em]\scriptsize#1\end{tabular}}

\begin{document}
\title{Solving the 2D Advection-Diffusion Equation using Fixed-Depth Symbolic Regression and Symbolic Differentiation without Expression Trees}
%
%
\author{Edward Finkelstein\inst{1}
}
\authorrunning{Edward F.} 
%
\institute{SDSU, USA\\
\email{efinkelstein5749@sdsu.edu}}
\maketitle              
\begin{abstract}
This paper presents a novel method for solving the 2D advection-diffusion equation using fixed-depth symbolic regression and symbolic differentiation without expression trees. The method is applied to two cases with distinct velocity fields and initial- and boundary-condition constraints. This method offers a promising solution for finding approximate analytical solutions to differential equations, with the potential for application to more complex systems involving vector-valued objectives.
\end{abstract}

\section{Introduction}
\subsection{Symbolic Regression}
Symbolic regression (SR) is a method that seeks a symbolic model $f\left(\vec{x}\right)$ to predict a (typically scalar) label $y$ based on an N-dimensional feature vector $\vec{x}$ which minimizes a loss metric $\mathcal{L}\left(f\left(\vec{x}\right),y\right)$ \cite{radwan2024comparisonrecentalgorithmssymbolic}. The nodes of the expression $f$ usually fall into one of the following three categories \cite{lacava2021contemporarysymbolicregressionmethods}:
\begin{itemize}
\item \textbf{Unary operators: } Any operator of arity 1, such as $\cos$, $\sin$, $\exp$, $\ln$, $\tanh$, etc.
\item \textbf{Binary operators: } Any operator of arity 2, such as $+$, $-$, $*$, $\div$, etc.
\item \textbf{Leaf Nodes: } Any of the individual features $\vec{x} = \{x_1, x_2, \ldots,x_{N}\}$ and constant tokens that can potentially be optimized with non-linear optimization routines such as L-BFGS \cite{doi:10.1137/0916069} or Levenberg-Marquardt \cite{83b09f23-b20e-3617-8f72-24765b713f7b} \cite{doi:10.1137/0111030}.
\end{itemize}
The models distilled by symbolic regression are inherently interpretable compared to other machine learning methods, which is advantageous in many disciplines \cite{lacava2021contemporarysymbolicregressionmethods} such as health, law, the natural sciences, etc. 
In the context of differential equations, the optimization problem can be framed as finding a model $\hat{f}$ such that 
\begin{equation}
\hat{f} = \textunderset{$\hat{f} \in \mathcal{F}$}{\text{argmin}} \; \mathcal{D}(\hat{f}) \;\text{  subject to  }\; \mathcal{I}, \vec{\mathcal{B}}
\end{equation}
where $\mathcal{F}$ is the space of all possible symbolic models under consideration, $\mathcal{D}$ is a differential operator representing the differential equation $\mathcal{D}(f) = 0$ ($f$ being the true model), $\,\mathcal{I}$ is the initial condition, and $\vec{\mathcal{B}}$ are the boundary conditions.
\subsection{Symbolic Differentiation}
Symbolic differentiation is a computer algebra approach to solving derivatives analytically \cite{10.1007/978-3-642-57201-2_12} through techniques such as the chain, product, and quotient rules. Symbolic differentiation falls under the umbrella of ``computer algebra'' technology \cite{tan2000symbolicc}, which modern-day software such as Mathematica \cite{Mathematica}, SymPy \cite{10.7717/peerj-cs.103}, Maxima \cite{maxima}, Maple \cite{maple}, and many others have made generally available. However, as remarked in \cite{PredragV2001}, many symbolic differentiation software rely on creating tree or graph data structures for the input formulae and the corresponding derivatives. While these data structure enables versatility, they may not be ideal for symbolic regression, which requires minimal memory allocations and maximal speed for optimal performance \cite{virgolin2022symbolicregressionnphard}. 
\par Thus, in this paper, the symbolic differentiation method detailed in \cite{PredragV2001}, as well as the in-situ simplification routines detailed therein, are implemented. Furthermore, the analogous differentiation techniques of \cite{PredragV2001} using Polish notation (prefix) expressions (as opposed to Reverse Polish notation (postfix)) are also implemented herein. The array-based method of \cite{PredragV2001} reduces memory allocation cost as no reallocations are needed for creating expression trees, and, since expressions are represented in this paper using prefix/postfix notation (using the grammars and algorithms developed in \cite{finkelstein2024generalizedfixeddepthprefixpostfix}), the conversion from infix to prefix/postfix is also eliminated, making the method developed in this paper highly efficient.

\subsection{2D Advection Diffusion Equation}
The two-dimensional advection-diffusion equation: 
\begin{align}
    &\frac{\partial T}{\partial t} + \nabla\cdot\left(\vec{u} T\right) = \kappa \nabla\cdot\left(\nabla T\right) \nonumber \\ 
    &\frac{\partial T}{\partial t} + \left(\frac{\partial}{\partial x} \hat{x} + \frac{\partial}{\partial y}\hat{y}\right)\cdot \left(T u_x \hat{x} + T u_y \hat{y}\right) = \kappa \left(\frac{\partial}{\partial x} \hat{x} + \frac{\partial}{\partial y}\hat{y}\right)\cdot \left(\frac{\partial T}{\partial x}\hat{x} + \frac{\partial T}{\partial y}\hat{y}\right) \nonumber \\
    &\frac{\partial T}{\partial t} + \frac{\partial}{\partial x}\left(T u_x\right) + \frac{\partial}{\partial y}\left(T u_y\right) = \kappa \left(\frac{\partial^2 T}{\partial x^2} + \frac{\partial^2 T}{\partial y^2}\right) \nonumber \\
    &\frac{\partial T}{\partial t} + \frac{\partial}{\partial x}\left(T u_x\right) + \frac{\partial}{\partial y}\left(T u_y\right) - \kappa \left(\frac{\partial^2 T}{\partial x^2} + \frac{\partial^2 T}{\partial y^2}\right) = 0
    \label{eq:AdvectionDiffusion2D},
\end{align}

is frequently used to model heat transfer and diffusion, as well as the movement of gases and fluids through various media \cite{AdvectionDiffusion2D}. In the case of constant $\vec{u}$ and $\kappa$, analytical solutions are known \cite{AdvectionDiffusion2D} \cite{wbb14}. In this work, two cases of constant $\kappa$ and $u_x$, $u_y$ constant in $x$, $y$, respectively, are considered, which means that equation \ref{eq:AdvectionDiffusion2D} can be rewritten as 

\begin{equation}
    \frac{\partial T}{\partial t} + u_x\frac{\partial T}{\partial x} + u_y\frac{\partial T}{\partial y} - \kappa \left(\frac{\partial^2 T}{\partial x^2} + \frac{\partial^2 T}{\partial y^2}\right) = 0 \label{eq:AdvectionDiffusion2D_Simplified}
\end{equation}

In this work, approximate analytical equations for $T$ are sought such that \ref{eq:AdvectionDiffusion2D_Simplified}, subject to two different $\mathcal{I}$ and $\vec{\mathcal{B}}$ is approximately satisfied.

\section{Related Work}
In \cite{oh2023geneticprogrammingbasedsymbolic}, the Euler-Bernoulli and Poisson equations subject to boundary constraints are solved using the Bingo framework \cite{10.1145/3520304.3534031} with automatic differentiation done in Pytorch. \cite{paszke2019pytorchimperativestylehighperformance}. The authors of \cite{oh2023geneticprogrammingbasedsymbolic} do not consider initial condition constraints, and they remark that differentiation with Pytorch performed sub-optimally compared to direct function evaluation.
\par The authors of \cite{Manti_2024} solve Poisson, Euler Elastica, and linear elasticity boundary-value problems using a custom library of discrete-exterior calculus operators implemented using the deap \cite{10.5555/2503308.2503311}, Ray \cite{10.5555/3291168.3291210}, Jax \cite{jax2018github} and pygmo \cite{Biscani2020} libraries for strongly typed genetic programming, parallelization, automatic differentiation, and LBFGS optimization, respectively. However, the framework \cite{Manti_2024}'s ability to work with dynamical systems is stated to be forthcoming \cite{desilva2020}.

\par In \cite{Kaptanoglu2022}, the assumptions is made that the governing equation $f$ in the dynamical system $d\vec{x}(t)/dt = \vec{f}(\vec{x}(t))$ is of the form $\sum_{k=1}^{p} \theta_{k}(\vec{x}) \xi_{k}$, where $\theta_k$ are pre-specified basis functions and $\xi_{k}$ the corresponding coefficients. Due to the sparsity assumption, \cite{desilva2020}, \cite{Kaptanoglu2022}, PySindy can obtain $f$ relatively quickly compared to traditional symbolic regression methods. However, the results can vary greatly depending on the choice of basis functions.
\par Lastly, \cite{sun2023symbolicphysicslearnerdiscovering} introduces the Symbolic Physics Learner (SPL) machine, which is tested on chaotic systems such as the double pendulum and the 3-dimensional Lorenz system. \cite{sun2023symbolicphysicslearnerdiscovering}'s acknowledged limitations, specifically the lack of multi-threading and the inaccuracy in approximating state derivatives, are addressed in this work by implementing a multi-threaded C++ Eigen \cite{eigenweb} symbolic regression framework, which leverages the symbolic differentiation method of \cite{PredragV2001}, ensuring exact derivative computation and significantly improved efficiency. 
\par The repository branch for this paper can be found \href{https://github.com/edfink234/Alpha-Zero-Symbolic-Regression/tree/PrefixPostfixSymbolicDifferentiator}{here}.
\section{Methods}
Expressions are generated using the framework for fixed-depth grammar algorithms developed in \cite{finkelstein2024generalizedfixeddepthprefixpostfix} for prefix/postfix notation expressions. The algorithms implemented therein, namely, Random Search, Monte-Carlo Tree Search, Particle Swarm Optimization, Genetic Programming, and Simulated Annealing, are extended here to support multi-threading such that each thread runs a separate version of the search algorithm. A concurrent map (using Boost's concurrent flat map API \cite{BoostLibrary}) containing the fitted constants of each unique expression encountered is shared. 
\par In addition, another less-trivial version of multi-threaded Monte-Carlo Tree Search, termed ``Concurrent MCTS'', is implemented here such that, in addition to the map of shared fitted constants, three additional concurrent maps are shared for storing the number of times expression $s$ was visited, $N(s)$, the number of times token $a$ was selected from state $s$, $N(s, a)$, and the estimated reward of taking action $a$ from state $s$, $Q(s, a)$. The MCTS implementation of \cite{finkelstein2024generalizedfixeddepthprefixpostfix} is slightly modified such that a \emph{random} token $a$ is selected from state $s$ in the case that multiple possible tokens from the current expression haven't been visited, i.e., $N(s,\vec{a}) = 0$. This modification ensures that multiple threads aren't searching the exact same branches of the search space.
\par Once an expression $f$ is generated, it is plugged into the differential equation to produce another expression which is then evaluated on the considered domain to compute the residual. Exact derivatives are computed symbolically using the Reverse Polish Notation method developed in \cite{PredragV2001} for postfix expressions and the analogous Polish Notation method developed herein for prefix expressions. 
 
\section{Experiments and Results}
This section considers two cases of the advection-diffusion equation with different $u_x$, $u_y$ and initial/boundary conditions. Initial tests are performed to determine the best-suited algorithm (Random Search, Monte-Carlo Tree Search, Concurrent Monte-Carlo Tree Search, Particle Swarm Optimization, Genetic Programming, and Simulated Annealing), expression depth (depth of expression to be generated), expression representation (prefix/postfix), and which constant tokens to include in the search space, namely if either \textbf{1.)} only the input variables should be considered, \textbf{2.)} if both input variables and constant tokens (not to be optimized) should be considered, or if \textbf{3.)} input variables, non-optimizable, \emph{and} optimizable constants should be considered\footnote{See \cite{finkelstein2024generalizedfixeddepthprefixpostfix} for details about the configurations tested herein.}. Then, the search is performed with the intent of minimizing the objective:
\begin{equation}
    \mathrm{MSE}_{\mathrm{tot}} = \mathrm{MSE}_{\mathcal{D}} + \sum_i \mathrm{MSE}_{\mathcal{B}_{i}} + \mathrm{MSE}_{\mathcal{I}},
\end{equation}
where satisfying the differential equation $\mathcal{D}$, each of the boundary conditions $\mathcal{B}_i$, and the initial condition $\mathcal{I}$ are weighted equally.
\par The considered unary operators for both cases are \textbf{-}, $\mathbf{log}$, $\mathbf{exp}$, $\mathbf{cos}$, $\mathbf{sin}$, $\mathbf{sqrt}$, $\mathbf{asin}$, $\mathbf{acos}$, $\mathbf{tanh}$, $\mathbf{sech}$. The considered binary operators are \textbf{+}, \textbf{-}, \textbf{*}, \textbf{/}, $\bm{\wedge}$. The considered non-optimizable constant tokens are \textbf{0}, \textbf{1}, \textbf{2}, \textbf{4}, and the minimum and maximum values for each input variable. For each case, the initial condition is stored as an attribute of the dataset\footnote{Thanks to Dr. Hamed Masnadi Shirazi for this idea.} such that the initial condition may be selected by the symbolic regressor, for which the corresponding derivatives wrt. $x$, $y$, and $t$ are evaluated accordingly.
\par Inspecting equation \ref{eq:AdvectionDiffusion2D_Simplified}, one may notice that simply setting $T=0$, or anything independent of $x$, $y$, and $t$, is an exact solution. To avoid this outcome, it is initially required that the derivative of $T$ with respect to each of the input variables is nonzero up to a threshold, which is set to $\mathbf{1/\sqrt{2}}$ for the experiments in this section. If a generated expression does not satisfy this constraint, the mean-squared error is artificially set to $\infty$ to discourage trivial expressions.
\par Lastly, in each of the two cases considered, 10 linearly spaced $x$, $y$, and $t$ values are sampled along the respective domains such that mesh grids of 1000 points are generated.
\subsection{Initial Tests}\label{subsec:InitTests1}
The following list denotes the configurations that were tested for the two cases of equation \ref{eq:AdvectionDiffusion2D_Simplified} considered in this paper\footnote{The test files ran for this subsection can be found \href{https://github.com/edfink234/Alpha-Zero-Symbolic-Regression/tree/708d1f2a774a0207da72c17a2626b10fff727e74/AdvectionDiffusionTests}{here}.}.
\begin{itemize}
    \item \textbf{Algorithms: } 
    \begin{enumerate}
        \item Random Search
        \item Monte Carlo Tree Search
        \item Concurrent Monte Carlo Tree Search
        \item Particle Swarm Optimization
        \item Genetic Programming
        \item Simulated Annealing
    \end{enumerate}
    \item \textbf{Expression Notations: } 
    \begin{enumerate}
        \item Polish Notation (prefix)
        \item Reverse Polish Notation (postfix)
    \end{enumerate}
    \item \textbf{Expression Depths: } $1 \leq N \leq 30$, where $N$ is the depth of the expression tree, using the convention that depth-0 expression trees correspond to leaf nodes
    \item \textbf{Constant Tokens in Basis Set: }
    \begin{enumerate}
        \item No constant tokens
        \item Non-optimizable tokens
        \item Non-optimizable tokens + optimizable token
    \end{enumerate}
\end{itemize}
Thus, $6 \times 2 \times 30 \times 3 = 1080$ configurations are tested for each case. Due to limited computational resources, each configuration is run for 5 seconds\footnote{This initial test is intended to suggest potentially suitable configurations, rather than being a definitive determination.}. The five best configurations obtained for Cases 1 and 2 are tabulated in Tables \ref{tab:Best_5_Case_1} and \ref{tab:Best_5_Case_2}, respectively. Additionally, the hyperparameters for each algorithm tested in this section are tabulated below\footnote{Based on hyperparameter tuning and literature review.}:
\begin{itemize}
    \item \textbf{Random Search: } N/A
    \item \textbf{Monte Carlo Tree Search: } Exactly as prescribed in section 3.2 of \cite{finkelstein2024generalizedfixeddepthprefixpostfix}, except that the number of iterations before updating $c$ if the MSE hasn't decreased, $N_{\mathrm{iter}}$, is \texttt{\textbf{500}} instead of 50,000.
    \item \textbf{Concurrent Monte Carlo Tree Search: } 
    \begin{itemize}
        \item Initial value of Exploration Parameter $c$: \texttt{\textbf{1.4}}
        \item Number of iterations before updating $c$ if the MSE hasn't decreased, $N_{\mathrm{iter}}$: \texttt{\textbf{1000}}
        \item Amount to increase $c$ after $N_{\mathrm{iter}}$ iterations if the MSE hasn't decreased: \texttt{\textbf{1.4}}
        \item Value to reset $c$ to if the MSE has decreased: \texttt{\textbf{1.4}}
    \end{itemize}
    \item \textbf{Particle Swarm Optimization: } Exactly as prescribed in section 3.3 of \cite{finkelstein2024generalizedfixeddepthprefixpostfix}
    \item \textbf{Genetic Programming: } Exactly as prescribed in section 3.4 of \cite{finkelstein2024generalizedfixeddepthprefixpostfix}, except that the initial population is \texttt{\textbf{200}} instead of 2000 and the total population after crossover/mutation is $\approx$ \texttt{\textbf{400}} instead of $\approx$ 4000.
    \item \textbf{Simulated Annealing: } Exactly as prescribed in Section 3.5 of \cite{finkelstein2024generalizedfixeddepthprefixpostfix}
    
\end{itemize}

Each configuration was run using 8 threads on a single MacBook Pro with an M1 Core and $\sim$ 16 GB of usable RAM. Optimizable tokens, if added to the expression being built, were optimized using 5 iterations of Particle-Swarm Optimization, due to the nontriviality of embedding initial/boundary constraints into the loss function for C++ nonlinear optimization libraries like \href{https://github.com/yixuan/LBFGSpp}{LBFGSpp} and Eigen. Additionally, for cases like the Advection-Diffusion Equation, it is possible for constant optimization to result in $T$ becoming constant in any of its dependent variables, even if not so originally. Thus, constant optimization can be problematic in this case as it requires a way to regularize the loss function, perhaps by adding a penalty term proportional to the ``constness'' of $T$.

\begin{table}
\caption{Best 5 Configurations for Case 1 (Section \ref{subsec:case_1})}
\label{tab:Best_5_Case_1}
\begin{tabular}{llrlrrr}
\hline
\textbf{\#} & \textbf{Algorithm} & \textbf{Depth} & \textbf{Notation} & \textbf{MSE} & \textbf{Non-Optimizable Tokens} & \textbf{Optimizable Token} \\
\hline
1 & PSO & 16 & postfix & 1.391380 & \texttt{True} & \texttt{False} \\
2 & GP & 4 & postfix & 1.857070 & \texttt{False} & \texttt{False} \\
3 & GP & 6 & postfix & 2.328540 & \texttt{True} & \texttt{False} \\
4 & MCTS & 14 & postfix & 2.416220 & \texttt{True} & \texttt{False} \\
5 & Simulated Annealing & 12 & postfix & 2.512110 & \texttt{False} & \texttt{False} \\
\bottomrule
\end{tabular}
\end{table}

\begin{table}
\caption{Best 5 Configurations for Case 2 (Section \ref{subsec:case_2})}
\label{tab:Best_5_Case_2}
\begin{tabular}{llrlrrr}
\hline
\textbf{\#} & \textbf{Algorithm} & \textbf{Depth} & \textbf{Notation} & \textbf{MSE} & \textbf{Non-Optimizable Tokens} & \textbf{Optimizable Token} \\
\hline
1 & Simulated Annealing & 8 & prefix & 12.664100 & \texttt{True} & \texttt{False} \\
2 & Random Search & 7 & postfix & 13.525700 & \texttt{True} & \texttt{False} \\
3 & MCTS & 13 & postfix & 13.528000 & \texttt{True} & \texttt{False} \\
4 & PSO & 9 & postfix & 22.711300 & \texttt{False} & \texttt{False} \\
5 & Simulated Annealing & 11 & prefix & 23.023000 & \texttt{True} & \texttt{False} \\
\bottomrule
\end{tabular}
\end{table}

\subsection{Case 1} \label{subsec:case_1}
In this case, equation \ref{eq:AdvectionDiffusion2D_Simplified} is considered for the following values:
\begin{align}
    u_x &= 1-y^2 \nonumber \\
    u_y &= 0 \nonumber \\
    \kappa &= 1 
\end{align}
The following domain is considered for the simulation:
\begin{equation}
     0.1 \leq x \leq 2.1, \quad -1.1 \leq y \leq 1.1, \quad 
     0.1 \leq t \leq 20
\end{equation}

\noindent The following boundary condition constraints are implemented in this case:
\begin{align}
    \frac{dT(y = -1.1)}{dy} &= \frac{dT(y = 1.1)}{dy} = 0 \nonumber \\
    T(x = 0.1) &= T(x = 2.1)  \nonumber \\
    \frac{dT(x = 0.1)}{dx} &= \frac{dT(x = 2.1)}{dx}
\end{align}
The following initial condition constraint is utilized:
\begin{equation}
    \mathcal{I} = \frac{\exp{\left(-((x - 1.1)^2 + y^2)\right)}}{0.08}
\end{equation}
\subsubsection{Results}\label{subsubsec:result_case_1}
\hspace{1cm} \\[0.2cm]
The configuration that resulted in the lowest Mean Squared Error (MSE) in this instance was configuration \#3. Although configuration \#1 was initially tested, it remained trapped in a local minimum for an extended period, yielding an MSE of 24.1156. Configuration \#2 exhibited improved performance with an MSE of 0.0365547 yet also converged to a local minimum. Anticipating that Configuration \#3 could potentially offer a more favorable bias-variance tradeoff, it was tested and yielded an \text{MSE = 0.00379332} before reaching an impasse. 
\par The lowest MSE equation obtained for this case before simplification (\ref{eq:Case1BestUnsimplified}) and after simplification (\ref{eq:Case1BestSimplified}) is as follows:
\begin{align}
    T &= \left( \mathcal{I}^{\tanh(\mathcal{I})^{\sqrt{t}}} \right) - \left( \operatorname{sech}\left( \mathcal{I} + \frac{t}{0.2 \cdot y} \right) \cdot \operatorname{sech}\left( x + \left( y + 2^{\mathcal{I}} \right) \right) \right) \label{eq:Case1BestUnsimplified}\\
    T &= \left( \mathcal{I}^{\tanh(\mathcal{I})^{\sqrt{t}}} \right) - \left( \operatorname{sech}\left( \mathcal{I} + \frac{t}{0.2 \cdot y} \right) \cdot \operatorname{sech}\left( x + y + 2^{\mathcal{I}} \right) \right)  \label{eq:Case1BestSimplified}
\end{align}









\subsection{Case 2} \label{subsec:case_2}
In this case, equation \ref{eq:AdvectionDiffusion2D_Simplified} is considered for the following values:
\begin{align}
    u_x &= \sin(4y) \nonumber \\
    u_y &= \cos(4x) \nonumber \\
    \kappa &= 1 
\end{align}
The following domain is considered for the simulation:
\begin{equation}
     0.1 \leq x \leq 2\pi, \quad 0.1 \leq y \leq 2\pi, \quad 
     0.1 \leq t \leq 20
\end{equation}

\noindent The following boundary condition constraints are implemented in this case:
\begin{align}
    T(x = 0.1) &= T(x = 2\pi) \nonumber \\
    \frac{dT(x = 0.1)}{dx} &= \frac{dT(x = 2\pi)}{dx}  \nonumber \\
    T(y = 0.1) &= T(y = 2\pi) \nonumber \\
    \frac{dT(y = 0.1)}{dy} &= \frac{dT(y = 2\pi)}{dy}  
\end{align}
The following initial condition constraint is utilized:
\begin{equation}
    \mathcal{I} = \frac{\exp{\left(-((x - \pi)^2 + (y - \pi)^2)\right)}}{0.08}
\end{equation}
\subsubsection{Results}\label{subsubsec:result_case_2}
\hspace{1cm} \\[0.2cm]

The configuration that resulted in the lowest Mean Squared Error (MSE) in this instance was configuration \#2. Configuration \#1 was tested first, but after some time, it remained trapped in a local minimum for an extended period, yielding an MSE of 0.784878. Configuration \#2 yielded the best result, an \text{MSE = 0.689072} before reaching an impasse. Configuration \#3 was tested several times and converged to a local minimum, with the best MSE achieved being 0.749346.
\par The lowest MSE equation obtained for this case before simplification (\ref{eq:Case2BestUnsimplified}) and after simplification (\ref{eq:Case2BestSimplified}) is as follows:

\begin{align}
    T &= \mathcal{I}^{\frac{0.2^{t}}{\text{sech}(\log(\log(\pi))) + 0.2 + \frac{0.000000}{\left(\frac{x}{y + 0.100000}\right)}}} \label{eq:Case2BestUnsimplified} \\
    T &= \mathcal{I}^{\frac{0.2^{t}}{\text{sech}(\log(\log(\pi))) + 0.2 }} \approx \mathcal{I}^{0.2^t / 1.191} \label{eq:Case2BestSimplified}
\end{align}









\subsubsection{Improved Results}\label{subsubsec:result_case_2_improved}
\par \hspace{1cm} \\ 
The relatively large mean-squared error achieved for Case 2 in section \ref{subsubsec:result_case_2} compared to the mean-squared error achieved for Case 1 in section \ref{subsubsec:result_case_1} prompted further investigation into this case. Upon inspecting the corresponding expression tree for equation \ref{eq:Case2BestSimplified}, it was observed that the expression could have been represented using a shallower tree. Thus, a depth of 5 was used in this section instead of 7, and the search was run again using GP (based on passed tests that had shown this algorithm to work well in practice). This search resulted in an improved \text{MSE = 0.0117411} and the following expression before \ref{eq:Case2ImprovedUnsimplified} and after \ref{eq:Case2ImprovedSimplified} 
 simplification.

\begin{align}
    T &= \left( \mathcal{I}^{1^{\arcsin\left(\sqrt{\mathcal{I}}\right)}} \right) + \operatorname{sech}\left( \frac{\left( \frac{2}{6.283185} + (x + y) \right)}{(t \cdot 20) \cdot (t + 0)} \right) \label{eq:Case2ImprovedUnsimplified} \\
    T &= \mathcal{I} + \operatorname{sech}\left( \frac{1/\pi + x + y}{20t^2} \right)
\label{eq:Case2ImprovedSimplified} 
\end{align}

\subsection{Simulated Annealing of Best Results}
To improve the results obtained in sections \ref{subsubsec:result_case_1} and \ref{subsubsec:result_case_2_improved}, the best-obtained expressions, namely, equations \ref{eq:Case1BestSimplified} and \ref{eq:Case2ImprovedSimplified} for cases 1 and 2 respectively, were used as the starting expressions for the Simulated Annealing algorithm. Additionally, Particle Swarm Optimization was also enabled in this section, and, rather courageously, the constant threshold was reduced from $\mathbf{1/\sqrt{2}}$ to $\mathbf{0}$. Finally, to increase confidence in the results, the resulting expressions were annealed on $500 \times 500 \times 500$ mesh grids (as opposed to the $10 \times 10 \times 10$ mesh grids used previously).
\subsubsection{Case 1}
Using equation \ref{eq:Case1BestUnsimplified} as the starting point of this procedure, a new expression was found which, to floating-point (4-byte) precision, yielded an \textbf{MSE = $\mathbf{0}$} (previously $3.79332 \times 10^{-3}$), the resulting expression of which is given in equations \ref{eq:Case1_annealed_unsimplified} (un-simplified) and \ref{eq:Case1_annealed_simplified} (simplified).

\begin{align}
    T &= \left(\mathcal{I}^{\operatorname{sech}(y_0)^{\arcsin(0.1)}}\right) - \left(\operatorname{sech}\left((y - x) - \frac{\operatorname{sech}(0.103287)}{0.1^{\mathcal{I}}}\right) \cdot \frac{\frac{t}{\sin(\operatorname{sech}(1.1))}}{t - 0.1^{0.1 + 1.1}}\right) \label{eq:Case1_annealed_unsimplified} \\
    T &= \left(\mathcal{I}^{\operatorname{sech}(y_0)^{\arcsin(0.1)}}\right) - \left(\operatorname{sech}\left((y - x) - \frac{\operatorname{sech}(0.103287)}{0.1^{\mathcal{I}}}\right) \cdot \frac{\frac{t}{\sin(\operatorname{sech}(1.1))}}{t - 0.1^{1.2}}\right) \label{eq:Case1_annealed_simplified}
\end{align}

\subsubsection{Case 2}
Employing the procedure in this section, equation \ref{eq:Case2ImprovedUnsimplified} was improved to yield an \textbf{MSE = $\mathbf{1.19209 \times 10^{-7}}$} (previously $1.17411 \times 10^{-2}$), the resulting expression of which is given in equations \ref{eq:Case2_annealed_unsimplified} (un-simplified) and \ref{eq:Case2_annealed_simplified} (simplified).

\begin{align}
    T &= \left( \mathcal{I}^{y_0^{\log \left(\frac{6.283185}{6.283185}\right)}} \right) + \operatorname{sech} \left( \frac{\arccos(0.819757) + (x + y)}{t \cdot 20.0 \cdot (12.499170)^2} \right) \label{eq:Case2_annealed_unsimplified} \\
    T &= \mathcal{I} + \operatorname{sech} \left( \frac{\arccos(0.819757) + (x + y)}{t \cdot 20.0 \cdot (12.499170)^2} \right) \label{eq:Case2_annealed_simplified}
\end{align}

\section{Remarks and Conclusions}
\par In comparing Case 1 and Case 2, the simpler Case 1 led to a lower mean squared error (MSE). Case 2's more complex velocity profile $\vec{u}$ and periodic boundary conditions resulted in a sparser solution space and costlier expression evaluations. For both cases, different algorithms showed varying success.
\par This study demonstrated an encourageable approach to finding approximate solutions to the 2D advection-diffusion equation for two cases of initial and boundary constraints by employing fixed-depth symbolic regression and symbolic differentiation without expression trees. The methods utilized proved very efficient due to the elimination of expression trees and constraining the search to fixed-depth prefix/postfix symbolic expressions. The results of the two test cases showcased the ability of this approach to handle the respective boundary and initial conditions, though they could be extended to include further constraints to constrict the space of solutions allowed by physics.

This work lays the foundation for further exploration in applying symbolic regression and differentiation to more complex differential equations and systems involving vector-valued objectives, such as the Navier-Stokes equations. Future work could explore alternative search algorithms to improve convergence rates and minimize local minimum entrapments within this framework.
\par Lastly, one potentially promising direction could be to develop a completely mesh-independent method of finding solutions to dynamical systems. This venture would include employing \emph{symbolic simplification} to the expression $\mathcal{D}(\hat{f})$. In this scenario, an exact solution would yield zero terms on the left hand side, and an approximate solution would yield a left-hand-side small in the number of terms and estimated magnitude.

\newpage
\bibliographystyle{splncs04}
\bibliography{PrefixPostfixPaper}


\end{document}